\documentclass{ws-mpla}
\usepackage[super]{cite}
\usepackage{graphicx}

\usepackage{slashed}
\usepackage{listings}
\usepackage{widetext}
\usepackage{graphics,bm}
\usepackage{epsfig}
\usepackage{graphicx}
\usepackage{amsmath}
\usepackage{amssymb}
\usepackage{bbold}
\usepackage{wrapfig,widetext}
\usepackage{color}
\usepackage{xcolor}
\usepackage{bbold}
\usepackage{hyperref}
\usepackage{mathtools}
\usepackage{cancel}
\usepackage{multirow}
\usepackage{upgreek}
\usepackage{subcaption}
\usepackage{marginnote}
    \usepackage[vertfit]{breakurl} 

\DeclareMathOperator{\Tr}{Tr}
\DeclareMathOperator{\z}{\mathfrak{z}}
\DeclareMathOperator{\bk}{\mathbf{k}}
\DeclareMathOperator{\bzero}{\mathbf{0}}

\begin{document}

\markboth{Prabal Adhikari, Elizabeth Leeser, Jake Markowski}{Phonon modes of magnetic vortex lattices in finite isospin chiral perturbation theory}

\catchline{}{}{}{}{}

\title{Phonon modes of magnetic vortex lattices in finite isospin chiral perturbation theory}

\author{Prabal Adhikari~\footnote{Corresponding author}}

\address{Physics Department,\\ Faculty of Natural Sciences and Mathematics,\\ St. Olaf College, 1520 St. Olaf Avenue,\\ Northfield, MN 55057, USA\\
adhika1@stolaf.edu}
\address{Kavli Institute for Theoretical Physics,\\ University of California,\\ Santa Barbara 93106,\\ USA}

\author{Elizabeth Leeser}
\address{St. Olaf College,\\ 1520 St. Olaf Avenue,\\ Northfield, MN 55057, USA\\leeser1@stolaf.edu}
\author{Jake Markowski}
\address{St. Olaf College,\\ 1520 St. Olaf Avenue,\\ Northfield, MN 55057, USA\\markow2@stolaf.edu}

\maketitle


\begin{abstract}
We study phonon modes associated with magnetic vortex lattices of finite isospin chiral perturbation theory near the upper critical point by introducing quasimomentum fluctuations to the lattice and calculate dispersion relations associated with the optical and acoustic modes. We find that one of the acoustic modes is massless and that its energy for small transverse quasimomentum is quartic (due the presence of an isospin chemical potential), which is significantly softer than the ``supersoft" (quadratic) massless mode of the Abelian Higgs Model (AHM). Due to the presence of derivative interactions, which is absent in the AHM, the speed of the longitudinal mode depends on both the isospin chemical potential and the external magnetic field.  Our results suggest that the standard assumption of an ordered lattice in finite isospin QCD should be revisited and the existence of a disordered spaghetti phase of a vortex liquid or gas, should be considered.

\end{abstract}


\section{Introduction}
The phase diagram of finite isospin Quantum Chromodynamics (QCD)~\cite{Son:2000xc,Son:2000by} has generated significant interest in recent years with potential applications to the early universe~\cite{Vovchenko:2020crk} and heavy ion collisions, which is isospin asymmetric. Furthermore, there are predictions of exotic astrophysical objects such as pion stars~\cite{Carignano:2016lxe,brandt2018new,Andersen:2018nzq} that are expected to form in supernovae remnants with the high neutrino density being the driving force that leads to the condensation of pions~\cite{Abuki:2009hx} in compact objects. Finite isospin QCD has the advantage of being accessible analytically at both low and asymptotically large isospin densities. At low densities, chiral perturbation theory ($\chi$PT)~\cite{Son:2000xc,Son:2000by} or equivalently the non-linear sigma model~\cite{Dashen:1974pio} shows the existence of a Bose-Einstein condensate and a superfluid. On the other hand, at asymptotically large isospin densities, the system has been shown to support a Bardeen-Cooper-Schrieffer (BCS) phase using high density effective theory~\cite{Kanazawa:2011tt,Kanazawa:2014lga,Cohen:2015soa} with the results being model-independent. Since finite isospin lattice QCD~\cite{Kogut:2002zg,Brandt:2017zck} does not suffer from a sign problem, intermediate densities have also been studied with results in the low and high density regimes consistent with analytical results with future searches focusing on a possible crossover transition from the Bose-Einstein condensate (BEC) to the BCS phase. For a review of finite isospin QCD, see Ref.~\cite{Mannarelli:2019hgn} and references therein including model dependent studies~\cite{He:2005sp,Adhikari:2018cea}.

The dynamics of the early universe and heavy ion collisions both occur in a magnetic field background suggesting that it is meaningful to explore the effect of the background on the phase diagram of finite isospin QCD. However, due to the asymmetry in the magnitude of the electromagnetic charges of the up and down quarks, the simultaneous presence of isospin and magnetic field introduces a sign problem to lattice QCD. Fortunately, due to the model-independent nature of $\chi$PT, it is possible to map out the phase diagram of QCD at finite isospin density and a magnetic background in a regime where the isospin chemical potential ($\mu_{I}$) and magnetic field ($\sqrt{eH}$) are both small compared to the typical hadronic scale ($\Lambda_{\rm Had}\sim4\pi f_{\pi}$), where $f_{\pi}$ is the pion decay constant. In particular, the charged pion superfluid that forms for isospin chemical potentials ($\mu_{I}$) larger than the pion mass ($m_{\pi}$) behaves as an extreme type-II superconductor in a magnetic background for realistic physical parameters with the phase diagram allowing for either a single vortex near a lower critical magnetic field ($H_{c1}$), a magnetic vortex lattice below an upper critical magnetic field ($H_{c2}$) or a normal phase above the upper critical field. For $\mu_I>m_{\pi}$, the system ssupports a superfluid pionic phase in the absence of a magnetic background~\cite{Adhikari:2015wva,Adhikari:2018fwm}. There is a further phase induced by the chiral anomaly at finite isospin $\chi$PT that supports a neutral pion soliton lattice~\cite{Brauner:2016pko,Brauner:2021sci}. A recent study~\cite{gronli2022competition} analyzed the competition between the magnetic vortex lattice phase with the chiral soliton lattice, which showed that for isospin chemical potentials less than than approximately $500\ {\rm MeV}$ and magnetic fields ($H$) less than $0.8\ {\rm GeV^{2}}$, the vortex lattice remains the ground state with the chiral soliton lattice generally favored for larger values of the isospin chemical potential and magnetic field, see Ref.~\cite{gronli2022competition} for the phase diagram.~\footnote{We note in passing an even more recent study showing a magnetic vortex lattice that supports a neutral pion current induced by the chiral anomaly at finite baryon density, see Ref.~\cite{Evans:2022hwr}. }

In this paper, we add to the emerging picture of the phase diagram of finite isospin QCD by studying the phonon modes associated with the triangular magnetic vortex lattice below and near the upper critical magnetic field~\cite{Adhikari:2018fwm}. In particular, we find the dispersion relations of the two acoustic and two optical modes by considering the quasimomentum fluctuations of the ground state in the lowest Landau level approximation (LLL). We find that one of the acoustic modes is massless with a dispersion relation that is quartic for small quasimomenta. We discuss the implications of this result for the phase diagram of finite isospin QCD including the possibility of magnetic vortices existing in a disordered (liquid/gas) phase. 

The paper is organized as follows: we begin with a brief review of pion condensation in finite isospin $\chi$PT in Section~\ref{sec:Lagrangian} followed by a review of the origin of superconductivity in Section~\ref{sec:superconductivity}. In Section~\ref{sec:phonon}, we introduce quasimomentum modes to the magnetic vortex solution in the LLL approximation. We then find the contribution to the potential and kinetic energies quadratic in the quasimomentum modes and find the dispersion relation of the optical and acoustic modes. Finally, we conclude in Section~\ref{sec:conclusion}, with a discussion of possible scenarios for the phase diagram of the finite isospin QCD supporting magnetic vortices existing not only in an ordered (solid) phase but also a disordered (liquid/gas) phase.

\section{Lagrangian and the origin of superconductivity}
\label{sec:Lagrangian}
\noindent
\subsection{Lagrangian}
We begin with the $\mathcal{O}(p^{2})$ $\chi$PT Lagrangian
\begin{align}
\label{eq:L2}
\mathcal{L}_{2}=&-\frac{1}{4}F_{\mu\nu}F^{\mu\nu}+\frac{f_{\pi}^{2}}{4}\Tr\left [\nabla_{\mu}\Sigma(\nabla^{\mu}\Sigma)^{\dagger} \right ]+\frac{f_{\pi}^{2}m_{\pi}^{2}}{4}\Tr\left[\Sigma+\Sigma^{\dagger}\right ]\ ,
\end{align}
where $\Sigma$ is an $SU(2)$ matrix, $f_{\pi}$ is the pion decay constant, $m_{\pi}$ is the pion mass, $F_{\mu\nu}\equiv\partial_{\mu}A_{\nu}-\partial_{\nu}A_{\mu}$ is the electromagnetic tensor (with $A_{\mu}$ the electromagnetic gauge potential) and $\nabla_{\mu}\Sigma$ is the covariant derivative,
\begin{align}
\nabla_{\mu}\Sigma&=\partial_{\mu}\Sigma-i[eQA_{\mu}+\tfrac{\mu_{I}}{2}\tau_{3}\delta_{\mu0},\Sigma]\ ,
\end{align}
where $\mu_{I}$ is the isospin chemical potential, $Q={\rm diag}\left(+\tfrac{2}{3},-\tfrac{1}{3}\right)$ is the quark charge matrix and we assume that $A_{0}(\infty)=0$, which ensures that the effect of the isospin chemical potential cannot be removed by a gauge transformation. The Lagrangian is $\mathcal{O}(p^{2})$ since the gauge field, $A_{\mu}$ and the $SU(2)$ field $\Sigma$ each count as $\mathcal{O}(p^{0})$ while derivatives ($\partial_{\mu}$), the isospin chemical potential ($\mu_{I}$), the electric charge ($e$) and pion mass ($m_{\pi}$) each count as $\mathcal{O}(p)$. Since $\chi$PT is an effective theory that encodes the interactions of the pseudo-Goldstone modes that arise in QCD due to chiral symmetry breaking, the results are strictly valid for scales significantly smaller than the typical hadronic scale of $4\pi f_{\pi}\sim 1\ {\tt GeV}$. As such, our analysis applies for $\mu_{I},\sqrt{eH}\ll 4\pi f_{\pi}$.

In two-flavor $\chi$PT, it is most convenient to parameterize the homogeneous ground state represented by the $SU(2)$ field, $\Sigma$, in terms of the Pauli matries, $\tau_{a}$ and unit vectors $\hat{\phi}_{a}$,
\begin{align}
\Sigma&=\cos\rho\ \mathbb{1}+\sin\rho\ i\hat{\phi}_{a}\tau_{a}\ .
\end{align}
We have adopted the Einstein summation convention over the repeated isospin index $a$ with $\hat{\phi_{a}}$ satisfying $\hat{\phi}_{a}\hat{\phi}_{a}=1$, which ensures that $\Sigma$ is unitary, i.e. $\Sigma^{\dagger}\Sigma=\mathbb{1}$. The resulting tree-level Lagrangian in the presence of an isospin chemical potential is 
\begin{align}
\mathcal{L}_{\rm tree}&=\frac{f_{\pi}^{2}m_{\pi}^{2}}{2}\left [2\cos\rho+\sin^{2}\rho\left\{\tfrac{\mu_{I}^{2}}{m_{\pi}^{2}}(\hat{\phi}_{1}^{2}+\hat{\phi}_{2}^{2})\right\} \right ]\ ,
\end{align}
which is independent of $\hat{\phi}_{3}$. Maximizing it then requires setting $\hat{\phi}_{1}^{2}+\hat{\phi}_{2}^{2}=1$ in the ground state and also implies that $\hat{\phi}_{3}=0$, i.e. the neutral pion doesn't condense. A further maximization of the Lagrangian with respect to $\rho$ results in a normal vacuum with $\rho=0$ for $|\mu_{I}|\le m_{\pi}$ and pion condensed phase with $\rho=\arccos\left(\tfrac{m_{\pi}^{2}}{\mu_{I}^{2}}\right)$ for $|\mu_{I}|\ge m_{\pi}$, which is also a charged superfluid. As has been in shown in Ref.~\cite{Son:2000xc}, the resulting dispersion relation is linear for small momenta.
\subsection{Origin of superconductivity}
\label{sec:superconductivity}
\noindent In the presence of a magnetic field, a charged superfluid behaves as a superconductor. In order to determine the nature of superconductivity (either type-I or type-II) in finite isospin $\chi$PT for isospin chemical potentials larger than the pion mass, the relevant thermodynamic quantity is the Gibbs free energy (density), 
\begin{align}
\mathcal{G}=\mathcal{H}-\vec{M}\cdot\vec{H}\ .
\end{align}
It is the Legendre transform of the Hamiltonian (density), $\mathcal{H}$ with $\vec{M}$ being the magnetization, which is defined in terms of the external magnetic field $\vec{H}$ and the magnetic field associated with the spatial components of the dynamical photon field $\vec{A}$, i.e.  $\vec{B}=\vec{\nabla}\times \vec{A}$, which is altered in the presence of the external magnetic field, 
\begin{align}
\vec{M}=\vec{B}-\vec{H}\ .
\end{align}
As is conventional, we assume $\vec{H}$ is uniform and points in the positive $z$-direction.

Assuming type-I superconductivity, we can determine the critical magnetic field by setting the Gibbs free energy of a uniform superconducting phase to that of the normal phase. In the superconductor, $\vec{B}=0$ or equivalently $\vec{M}=-\vec{H}$. We find
\begin{align}
H_{c}=\frac{f_{\pi}}{\mu_{I}}(\mu_{I}^{2}-m_{\pi}^{2})\ .
\end{align}
\begin{figure}[t!]
\begin{center}
\includegraphics[width=0.8\textwidth]{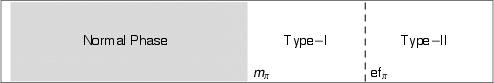}\\
\vspace{5pt}
\includegraphics[width=0.8\textwidth]{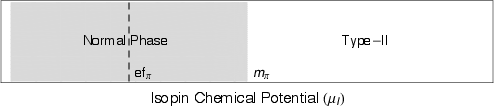}
\end{center}
\caption{Phase diagram illustrating the nature of superconductivity in finite isospin $\chi$PT. Top. $ef_{\pi}>m_{\pi}$: Both type-I and type-II superconductivity are supported for $\mu_{I}<ef_{\pi}$ and $\mu_{I}>ef_{\pi}$ respectively. Bottom. $ef_{\pi}<m_{\pi}$: Only type-II superconductivity is supported.} 
\label{fig:schematic}
\end{figure}

On the other hand, assuming type-II superconductivity, there are two critical magnetic fields, the lower one, $H_{c1}$, determines the transition from the uniform superconducting phase to a single vortex phase while the higher one, $H_{c2}$ determines the transition from a magnetic vortex lattice phase to the normal QCD vacuum. The lower critical field is characterized by the string tension, $\mathcal{S}$, which is the condensation energy per unit length of a single magnetic vortex relative to the superconducting phase
\begin{align}
H_{c1}=\frac{\mathcal{S}}{\Phi}\ ,
\end{align}
with $\Phi=\tfrac{2\pi}{e}$ being the lowest magnetic flux quantized by each vortex in the ground state.

As the external magnetic field is increased beyond $H_{c1}$ the ground state becomes populated with a finite density of magnetic vortices each with flux $\Phi$, though for large enough magnetic fields, magnetic vortex lattices become disfavored due to the large repulsive interactions that develop between vortices.  Beyond the upper critical field,
\begin{align}
\label{eq:Hc2}
H_{c2}\equiv B_{c}=\frac{\mu_{I}^{2}-m_{\pi}^{2}}{e}\ ,
\end{align}
the normal vacuum with no pion condensation ($\rho=0$) and zero magnetization ($\vec{M}=\vec{0}$) is the ground state. All of the external magnetic field, $\vec{H}$, penetrates the vacuum.

In Fig.~\ref{fig:schematic}, we illustrate the nature of superconductivity in finite isospin $\chi$PT. In order to understand the diagram, it is helpful to first note that $H_{c}$ under the assumption of type-I superconductivity and $H_{c2}$ under the assumption of type-II are proportional both to $\mu_{I}^{2}-m_{\pi}^{2}$. As such we define the ratio
\begin{align}
\mathcal{R}=\frac{H_{c2}}{H_{c}}&=\frac{\mu_{I}}{ef_{\pi}}\ ,
\end{align}
with the system behaving as a type-II superconductor if $H_{c2}\ge H_{c}$, i.e. $\mathcal{R}\ge1$ -- when the equality is satisfied then at the critical field the Gibbs free energy of a single vortex is equal to that of a vortex lattice suggesting that both can coexist. The condition $H_{c2}\ge H_{c}$, in combination with the requirement that $\mu_{I}>m_{\pi}$ for the superfluid phase, suggests that if $ef_{\pi}<m_{\pi}$, the system only supports type-II superconductivity -- this scenario is illustrated in the bottom diagram of Fig.~\ref{fig:schematic}. However, if $ef_{\pi}>m_{\pi}$, the system exhibits both type-I and type-II superconductivity for $\mu_{I}<ef_{\pi}$ and $\mu_{I}\ge ef_{\pi}$ respectively.

\section{Phonon modes}
\label{sec:phonon}
\noindent The key objective of the paper is to study the phonon modes that arise in magnetic vortices of the type-II superconductor in finite isospin $\chi$PT. However, prior to introducing quasimomentum fluctuations to the triangular magnetic vortex lattice that forms near the upper critical point and studying phonon dispersion relations, we briefly review the calculation of the vortex condensation energy first performed in Ref.~\cite{Adhikari:2018fwm}. For a perturbative calculation, it is useful to expand the $\chi$PT Lagrangian using the exponential~\footnote{In discussing the ground state of the $\vec{H}=0$ we used the linear parameterization of $\Sigma$, which is related to the exponential parametrization through $\rho=\frac{\sqrt{2\pi^{+}\pi^{-}+(\pi^{0})^{2}}}{f_{\pi}}$, $\hat{\phi}_{1}=\frac{1}{\sqrt{2}}\frac{\pi^{+}+\pi^{-}}{\sqrt{2\pi^{+}\pi^{-}+(\pi^{0})^{2}}}$, $\hat{\phi}_{2}=\frac{i}{\sqrt{2}}\frac{\pi^{+}-\pi^{-}}{\sqrt{2\pi^{+}\pi^{-}+(\pi^{0})^{2}}}$ and $\hat{\phi}_{3}=\frac{\pi^{0}}{\sqrt{2\pi^{+}\pi^{-}+(\pi^{0})^{2}}}$.} parameterization of $\Sigma$\ ,
\begin{align}
\Sigma&=\exp\left(i\frac{\Phi}{f_{\pi}}\right)\ ,&
\Phi&=\begin{pmatrix}\pi^{0}&\sqrt{2}\pi^{+}\\
\sqrt{2}\pi^{-}&-\pi^{0}
\end{pmatrix}\ ,
\end{align}
up to terms containing four pion fields,
\begin{align}
\nonumber
&\mathcal{L}_{4\pi}=-\tfrac{1}{4}F^{\mu\nu}F_{\mu\nu}-i\mu_{I}(\pi^{+}\partial_{0}\pi^{-}-\pi^{-}\partial_{0}\pi^{+})+2i\mu_{I}\ell^{2}\pi^{+}\pi^{-}(\pi^{+}\partial_{0}\pi^{-}-\pi^{-}\partial_{0}\pi^{+})\\ 
\nonumber
&+D_{\mu}\pi^{+}D^{\mu}\pi^{-}+(\mu_{I}^{2}-m_{\pi}^{2})\pi^{+}\pi^{-}-\tfrac{1}{2}\,\ell^{2}(4\mu_{I}^{2}-m_{\pi}^{2})(\pi^{+}\pi^{-})^{2}-\ell^{2}D_{\mu}\pi^{+}D^{\mu}\pi^{-}\pi^{+}\pi^{-}\\
&+\tfrac{1}{2}\ell^{2}\left[\pi^{+}D_{\mu}\pi^{-}\pi^{+}D^{\mu}\pi^{-}+D_{\mu}\pi^{+}\pi^{-}D^{\mu}\pi^{+}\pi^{-}\right]\ ,
\label{eq:L4}
\end{align}
where $\ell=\frac{1}{\sqrt{3}f_{\pi}}$ has been introduced for compactness, $\mu_{I}$ is the isospin chemical potential, $m_{\pi}$ is the pion mass, $f_{\pi}$ is the pion decay constant and the covariant derivatives associated with the pion fields are defined as $D_{\mu}\pi^{\pm}=(\partial_{\mu}+ieA_{\mu})\pi^{\pm}$. We exclude a trivial constant, $f_{\pi}^{2}m_{\pi}^{2}$ and a gauge-fixing term that is not necessary for a tree-level calculation.
\subsection{Method of successive approximation}
The analysis then proceeds using the method of successive approximation as in the seminal work by Abrikosov~\cite{Abrikosov:1956sx}. This allows for the calculation of the magnetic vortex lattice condensation energy and phonon modes of the magnetic vortex lattice. The technique amounts to utilizing the following power counting scheme
\begin{align}
B&=B_{0}+\epsilon^{2}\delta B\\
\pi^{+}&=\epsilon\tilde{\pi}^{+}+\epsilon^{3}\delta\pi^{+}\ ,
\end{align}
where at leading order $B_{0}$ is the critical magnetic field $B_{c}$ of Eq.~(\ref{eq:Hc2}) with $\tilde{\pi}^{+}$ obeying the following linearized equation of motion~\cite{Adhikari:2018fwm}
\begin{align}
\label{eq:pieom}
\left(2\bar{\partial}+\frac{eB_{0}}{2}z\right)\tilde{\pi}^{+}=0\ ,
\end{align}
where the use of ``conformal" coordinates,
\begin{align}
z&=x+iy\ ,\ \bar{z}=x-iy\\
\partial&=\tfrac{1}{2}(\partial_{x}-i\partial_{y}),\ \bar{\partial}=\tfrac{1}{2}(\partial_{x}+i\partial_{y})\ .
\end{align}
simplifies the analysis. The choice of power counting ensures that the leading order equation of motion is $\mathcal{O}(\epsilon)$ and the next-to-leading order equations of motion either contain $\delta B$ and two $\tilde{\pi}^{+}$ or $B_{0}$ and three $\delta\pi^{\pm}$ fields. The equation of motion for the magnetic field $B$ that incorporates next-to-leading order corrections is
\begin{align}
\nonumber
\bar{\partial}B&=-e\bar{\partial}(\tilde{\pi}^{+}\tilde{\pi}^{-})\left[1-2\ell^{2}\tilde{\pi}^{+}\tilde{\pi}^{-}\right]\ ,
\end{align}
which has the following solution that codifies magnetic flux conservation
\begin{align}
B&=B_{\rm ext}-e\left(\tilde{\pi}^{+}\tilde{\pi}^{-}-\langle\tilde{\pi}^{+}\tilde{\pi}^{-}\rangle\right)+e\,\ell^{2}\left((\tilde{\pi}^{+}\tilde{\pi}^{-})^{2}-\langle(\tilde{\pi}^{+}\tilde{\pi}^{-})^{2}\rangle\right)\ ,
\end{align}
with the external field, $H$, hitherto denoted $B_{\rm ext}$, $\langle\mathcal{Q}\rangle$ representing the expectation value of $\mathcal{Q}$ over the transverse plane with area $A_{\perp}$
\begin{align}
\label{eq:Q}
\langle\mathcal{Q}\rangle=\tfrac{1}{A_{\perp}}\int dx\, dy\ \mathcal{Q}\ .
\end{align}
The condensation energy, $\mathcal{E}_{\perp}$, of the magnetic vortex lattice, defined relative to the normal vacuum, 
\begin{align}
\mathcal{E}_{\perp}\equiv\langle\mathcal{H}\rangle-\tfrac{1}{2}B_{\rm ext}^{2}\ ,
\end{align}
can then be calculated using the solutions to the equations of motion~\cite{Adhikari:2018fwm} 
\begin{align}
\label{eq:E}
\mathcal{E}_{\perp}=-e(B_{c}-B_{\rm ext})\langle\tilde{\pi}^{+}\tilde{\pi}^{-}\rangle+\tfrac{e^{2}}{2}\langle\tilde{\pi}^{+}\tilde{\pi}^{-}\rangle^{2}+\tfrac{1}{2}\left(\lambda_{\rm eff}-e^{2}\right)\langle(\tilde{\pi}^{+}\tilde{\pi}^{-})^{2}\rangle\ ,
\end{align}
where $\lambda_{\rm eff}$ is defined as
\begin{align}
\label{eq:lambdaeff}
\lambda_{\rm eff}\equiv\tfrac{2\mu_{I}^{2}+m_{\pi}^{2}}{3f_{\pi}^{2}}\ .
\end{align}
It reduces to the repulsive quartic coupling of the AHM in the absence of derivative interactions, $\lambda_{\rm eff}\rightarrow \lambda_{\rm eff}(\ell=0)$
Restricting to the subspace of periodic lattices, the condensation energy is minimized by a triangular magnetic vortex lattice, for which the analytic form of the solution is
\begin{align}
\label{eq:pivortex}
\tilde{\pi}^{+}_{0}&=\sum_{n=-\infty}^{\infty}C_{n}\phi_{n}(\nu, z,\bar{z})\ ,&
\phi_{n}(\nu, z,\bar{z})&=e^{-\pi\nu^{2}n^{2}-\frac{\pi}{2L_{B}^{2}}(|z|^{2}+z^{2})+\frac{2\pi}{L_{B}}\nu nz}\ ,
\end{align}
where $L_{B}=\sqrt{\frac{2\pi}{eB}}$ is the magnetic length. Furthermore, for a triangular lattice that minimizes the free energy, $C_{n}=C_{n+2}$ with $C_{0}\equiv C$, $C_{1}=iC$ and $\nu=\frac{\sqrt[4]{3}}{\sqrt{2}}$. 

In order to calculate the transverse energy, we minimize Eq.~(\ref{eq:E}) by utilizing the definition of the Abrikosov ratio, $\beta_{A}=\frac{\langle(\tilde{\pi}^{+}\tilde{\pi}^{-})^{2}\rangle}{\langle\tilde{\pi}^{+}\tilde{\pi}^{-}\rangle^{2}}$, which is calculable for doubly periodic lattices. The resulting transverse energy in terms of the Abrikosov ratio is
\begin{align}
\mathcal{E}_{\perp}^{(0)}&=-\frac{e^{2}(B_{c}-B_{\rm ext})^{2}}{2\left[e^{2}+\beta_{A}\left (\lambda_{\rm eff}-e^{2}\right )\right]}\ ,
\end{align}
with subleading corrections of $\mathcal{O}\left ((B_{c}-B_{\rm ext})^{4} \right )$. The triangular lattice maximizes the Abrikosov ratio, consequently minimizing the free energy.
The resulting vortex lattice solution $\tilde{\pi}_{0}^{+}$ of Eq.~(\ref{eq:pivortex}) is then fully characterized by $|C|$, which depends on the $eB_{c}=\mu_{I}^{2}-m_{\pi}^{2}>0$,
\begin{align}
\label{eq:C}
|C|=\sqrt{\frac{1}{Q_{0}}\frac{e(B_{c}-B_{\rm ext})}{\beta_{A}(\lambda_{\rm eff}-e^{2})+e^{2}}}\ ,
\end{align}
with the Abrikosov ratio, $\beta_{A}=1.159595\cdots$ for the triangular lattice. $|C|$ vanishes for $B_{\rm ext}\ge B_{c}$ since magnetic vortices are not favored above the critical magnetic field of $B_{c}$.

\subsection{Quasimomentum Excitations}
A phonon mode in a magnetic vortex lattice, with energy $\omega$ and quasimomentum $\vec{k}$, introduces deformations that satisfy the equation of motion in Eq.~(\ref{eq:pieom}) and perturbs the lattice, whereby a mode with transverse quasimomentum, $\mathbf{k}=(k_{x},k_{y})$, distorts the lattice by $\frac{L_{B}^{2}}{\pi}(-k_{y},k_{x})$ where $L_{B}=\sqrt{\frac{2\pi}{eB}}$ is the magnetic length.~\cite{Chernodub:2014sia,Rosenstein:1999aa}. In the longitudinal ($\z$) direction, the quasimomentum excitations are plane waves
\begin{align}
\label{eq:quasimomentum-state}
\tilde{\pi}_{k}^{+}(\vec{x},t)=e^{-i(\vec{k}\cdot \vec{x}-\omega t) }\tilde{\pi}^{+}(\mathbf{x}+\tfrac{L_{B}^{2}}{\pi}\mathbf{\tilde{k}})\ ,
\end{align}
with $\vec{x}$ and $\mathbf{x}$ representing position and transverse position respectively -- explicitly, $\vec{x}=(\mathbf{x},\z)$ with $\mathbf{x}=(x,y)$ and $\vec{k}=(\bk,k_{\z})$ with $\mathbf{k}=(k_{x},k_{y})$ -- the transverse quasimomenta, $\mathbf{k}$, live within a hexagonal Brillouin zone~\footnote{see Fig.~2 in Ref.~\cite{Chernodub:2014sia}}. $\tilde{\mathbf{k}}$ is defined as $\mathbf{\tilde{k}}_{i}=\epsilon_{ij}\mathbf{k}_{j}$, where $\epsilon_{ij}$ is the antisymmetric tensor, $\epsilon_{12}=-\epsilon_{21}=1$. 

In order to incorporate the quasimomentum fluctuations of the magnetic vortex lattice, we expand $\tilde{\pi}^{\pm}$ around the triangular vortex lattice of Eq.~(\ref{eq:pivortex})~\cite{Rosenstein:1999aa} ,
\begin{align}
\label{eq:rep1}
\tilde{\pi}^{+}(\mathbf{x})&=\tilde{\pi}^{+}_{0}(\mathbf{x})+\chi^{\dagger}(\vec{x},t)\\
\label{eq:rep2}
\tilde{\pi}^{-}(\mathbf{x})&=\tilde{\pi}^{-}_{0}(\mathbf{x})+\chi(\vec{x},t)\ ,
\end{align}
with $\tilde{\pi}_{0}^{\pm}(\mathbf{x})$ representing the static triangular vortex lattice solution. $\chi^{\dagger}(\vec{x},t)$ and $\chi(\vec{x},t)$ represent perturbative fluctuations around the triangular vortex lattice 
\begin{align}
\label{eq:chi}
\chi^{\dagger}(\vec{x},t)=\sum_{\vec{k}\neq 0}c^{\dagger}_{k}\tilde{\pi}^{+}_{k}(\vec{x},t)\ ,
\end{align}
with the dimensionless coefficient, $c^{\dagger}_{k}$, being complex-valued and satisfying $|c^{\dagger}_{k}|\ll 1$, which ensures that the fluctuations are small. 
In considering the low-energy excitations of the magnetic vortex lattice, we have restricted to contributions arising through quasimomentum states in the lowest Landau level (LLL) -- this is justified as long as the quasimomentum is small compared to the gap to the excited Landau level, i.e. $|\vec{k}|\ll \sqrt{2eB_{\rm ext}}$.

In order to extract the phonon dispersion relations, we require the effective action associated with the phonons of the magnetic vortex lattice. It suffices to consider quasimomentum fluctuations that are quadratic in the ``operators" $c^{\dagger}_{k}$ and $c_{k}$, 
\begin{align}
\label{eq:S}
\mathcal{S}_{\rm phonon}^{(2)}=\sum_{\vec{k}\neq 0}\left(\mathcal{T}^{(2)}(\vec{k})-\mathcal{E}^{(2)}(\vec{k})\right)\ ,
\end{align}
where $\mathcal{T}^{(2)}(\vec{k})$ is the kinetic contribution arising from the time-dependent terms in Eq.~(\ref{eq:L4}) while $\mathcal{E}^{(2)}(\vec{k})$ is the potential energy contribution consisting of traverse and longitudinal contributions,
\begin{align}
\mathcal{E}^{(2)}(\vec{k})\equiv\mathcal{E}_{\perp}^{(2)}(\bk,0)+\mathcal{E}_{\parallel}^{(2)}(\mathbf{0},k_{\z})\ .
\end{align}
We will calculate the former using the transverse energy of Eq.~(\ref{eq:E}) while the latter requires utilizing the $\chi$PT Lagrangian of Eq.~(\ref{eq:L4}). The resulting quadratic contributions will contain the following two-point function
\begin{align}
\langle\tilde{\pi}^{+}_{\bk}\tilde{\pi}^{-}_{\mathbf{l}}\rangle&=|C|^{2}Q_{0}\delta(\bk-\mathbf{l})\ ,
\end{align}
where $Q_{0}\equiv\tfrac{1}{\sqrt{2}\nu}$ with $\nu=\frac{\sqrt[4]{3}}{\sqrt{2}}$, and $|C|$ determines the size of the pion condensate in the magnetic vortex lattice, see Eq.~(\ref{eq:C}), and vanishes at and above the critical magnetic field.Additionally, the following four-point function (with two non-zero quasimomenta) appear in the quadratic phonon action,
\begin{align}
\langle\tilde{\pi}^{+}_{\bk_{2}}\tilde{\pi}^{-}_{\mathbf{l}_{2}}\tilde{\pi}^{+}_{\bk_{1}}\tilde{\pi}^{-}_{\mathbf{l}_{1}}\rangle&\equiv|C|^{4}Q_{\bk_{2}\mathbf{l}_{2}\bk_{1}\mathbf{l}_{1}}\ .
\end{align}
$Q_{\bk_{2}\mathbf{l}_{2}\bk_{1}\mathbf{l}_{1}}$ is real, dimensionless and symmetric under the simultaneous quasimomenta exchange in the first and second indices, and the third and fourth indices,
\begin{align}
\label{eq:Qproperty}
Q_{\bk_{2}\mathbf{l}_{2}\bk_{1}\mathbf{l}_{1}}=Q_{\mathbf{l}_{2}\bk_{2}\mathbf{l}_{1}\bk_{1}}\ .
\end{align} 
Further details of the four-point function is relegated to \ref{sec:appendix} with detailed calculations found in Ref.~\cite{Chernodub:2014sia}. It is notable that the two-point function is non-vanishing for transverse momenta $\bk$ and $\mathbf{l}$ are equal -- this manifestation of phonon quasimomentum conservation will also be evident in the quadratic contributions that arise through four-pion field interactions.

The resulting quadratic phonon action, as will become evident from our calculations, is non-diagonal in the basis of $c_{k}$ and $c^{\dagger}_{k}$ -- this is not surprising since the quadratic action will contain contribution of the form $c_{k}c_{k}$ and $c^{\dagger}_{k}c^{\dagger}_{k}$. However, it  is diagonal in the real representation of $c^{\dagger}_{k}$, which is defined as
\begin{align}
\label{eq:c}
c^{\dagger}_{k}&=o^{\dagger}_{k}-ia^{\dagger}_{k}\ ,&\ c_{k}&=o_{k}+ia_{k}\ ,
\end{align}
where $a^{\dagger}_{k}$ and $o^{\dagger}_{k}$ represent excitations (``operators") associated with acoustic and optical phonon modes -- they satisfy $a^{\dagger}_{k}=a_{-k}$, $o^{\dagger}_{k}=o_{-k}$, and their quadratic forms are related to $c^{\dagger}_{k}$ and $c_{k}$ through the following identities
\begin{align}
c^{\dagger}_{k}c_{k}&=o^{2}_{k}+a^{2}_{k}\ ,&
\tfrac{1}{2}(c^{\dagger}_{k}c^{\dagger}_{-k}+c_{k}c_{-k})&=o^{2}_{k}-a^{2}_{k}\ .
\end{align}
The detailed calculation of the transverse, longitudinal and kinetic contribution to the phonon action requires, after the introduction of quasimomentum fluctuations an expansion in powers of the acoustic and optical phonon ``operators". Since the calculation is involved, we present the details in \ref{app:phononaction} while only stating the relevant results for the calculation of the phonon dispersion relation here. In particular, the quadratic kinetic, transverse and longitudinal contributions to the phonon action are
\begin{align}
\nonumber
\mathcal{T}^{(2)}(\vec{k})&=(\omega^{2}-2\mu_{I}\omega)|C|^{2}\left[Q_{0}-|C|^{2}\ell^{2}\Lambda_{+}(\bk)\right]o_{k}^{2}\\
\nonumber
&+(\omega^{2}-2\mu_{I}\omega)|C|^{2}\left[Q_{0}-|C|^{2}\ell^{2}\Lambda_{-}(\bk)\right]a_{k}^{2}\\
&+2\ell^{2}\mu_{I}\omega|C|^{4}(\tilde{\Lambda}_{+}(\bk)o_{k}^{2}+\tilde{\Lambda}_{-}(\bk)a_{k}^{2})\ ,\\
\nonumber
\mathcal{E}_{\perp}^{(2)}(\bk,0)&=\tfrac{1}{2}(\lambda_{\rm eff}-e^{2})|C|^{4}\left[\lambda_{-}(\bk)-\lambda_{-}(\bzero)\right]o_{\bk}^{2}\\
&+\tfrac{1}{2}(\lambda_{\rm eff}-e^{2})|C|^{4}\left[\lambda_{+}(\bk)-\lambda_{-}(\bzero)\right]a_{\bk}^{2}\ ,\\
\mathcal{E}_{\parallel}^{(2)}(\vec{k})&=k_{\z}^{2}|C|^{2}\left[Q_{0}-|C|^{2}\ell^{2}\Lambda_{+}(\bk)\right]o_{k}^{2}+k_{\z}^{2}|C|^{2}\left[Q_{0}-|C|^{2}\ell^{2}\Lambda_{-}(\bk)\right]a_{k}^{2}\ ,
\end{align}
where $\Lambda_{\pm}(\bk)$, $\tilde{\Lambda}_{\pm}(\bk)$ and $\lambda_{\pm}(\bk)$ are defined in terms of the four-point function as
\begin{align}
\Lambda_{\pm}(\bk)&=Q_{\bk\bk00}\pm Q_{\bk0-\bk0}\\
\tilde{\Lambda}_{\pm}(\bk)&=Q_{\bk\bk\bzero\bzero}+2Q_{\bk\bzero\bzero\bk}\pm Q_{\bk\bzero-\bk\bzero}\\
\lambda_{\pm}(\bk)&\equiv2(Q_{\bk\bk\bzero\bzero}+Q_{\bk\bzero\bzero\bk}\pm Q_{\bk\bzero-\bk\bzero})\ ,
\end{align}
with the low quasimomentum expansion of the four-point function discussed in \ref{sec:appendix}. As required for a quadratic action, two of the quasimomenta in the four-point function are zero with the explicit forms for small quasimomenta presented in Eqs.~(\ref{eq:Q1}) and (\ref{eq:Q2}). While the kinetic and longitudinal contributions depend on all components of the quasimomentum, the transverse contribution only depends on the transverse momentum. The longitudinal energy vanishes for zero longitudinal quasimomentum, $k_{\z}=0$ but also depends on the transverse quasimomentum due to the presence of derivative interactions ($\ell\neq0$).

\begin{figure*}[t!]
\centering
\includegraphics[width=0.45\textwidth]{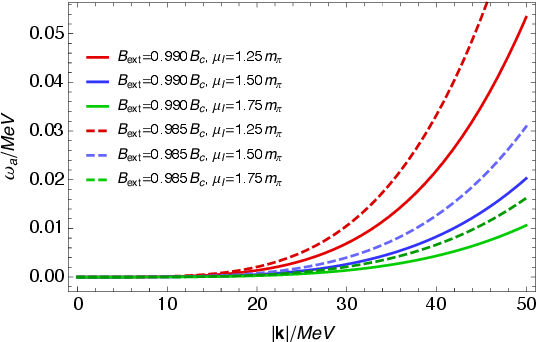}\,\,\,\,\,
\includegraphics[width=0.45\textwidth]{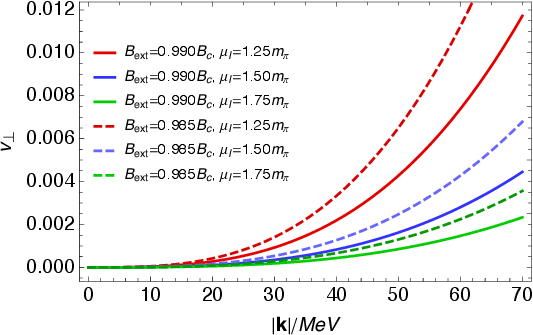}\,\,\,\,\,
\caption{Plot of the dispersion relation associated with the massless acoustic mode (left) and its transverse velocity (right) as function of the transverse quasimomentum, $|\bk|$. The solid and dashed lines represent the energy and speed at $B_{\rm }=0.99B_{c}$ and $B_{\rm }=0.985B_{c}$ respectively with the red, blue and green representing $\mu_{I}=1.25m_{\pi}$, $1.5m_{\pi}$ and $1.75m_{\pi}$ respectively.}  
\label{fig:dispersion}     
\end{figure*}
\subsection{Dispersion Relation}
\noindent
The kinetic contribution depends on the phonon energy, $\omega$, and the dispersion relation can be extracted from the effective action, Eq.~(\ref{eq:S}), through the equation of motion. This is equivalent to the condition $\mathcal{T}^{(2)}(k)-\mathcal{E}^{(2)}(k)=0$,
\begin{align}
(\omega^{2}-2\mu_{I}\omega)+2\ell^{2} \mu_{I}\omega g_{\pm}(\bk)=k_{\z}^{2}+\tilde{h}_{\pm}(\bk)\ ,
\end{align}
which has a compact expression due to following transverse quasimomentum functions 
\begin{align}
\tilde{h}_{\pm}(\bk)&=h_{\pm}(\bk)[\lambda_{\mp}(\bk)-\lambda_{-}(\bzero)]\\
g_{\pm}(\bk)&=\tfrac{|C|^{2}}{Q_{0}}\tilde{\Lambda}_{\pm}(\bk)\left(1+\tfrac{|C|^{2}\ell^{2}}{Q_{0}}\Lambda_{\pm}(\bk)\right)\\
\label{eq:hpm}
h_{\pm}(\bk)&=\tfrac{|C|^{2}}{2Q_{0}}(\lambda_{\rm eff}-e^{2})\left(1+\tfrac{|C|^{2}\ell^{2}}{Q_{0}}\Lambda_{\pm}(\bk)\right)\ .
\end{align}
For clarity, we briefly summarize the various quantities that enter the above results:  $|C|\sim \sqrt{e(B_{c}-B_{\rm ext})}$, of Eq.~(\ref{eq:C}), measures the size of the pion condensate while $Q_{0}=\frac{1}{\sqrt{2}\nu}$ with $\nu=\frac{\sqrt[4]{3}}{\sqrt{2}}$ for a triangular vortex lattice. $\ell=\frac{1}{\sqrt{3}f_{\pi}}$ characterizes the size of the derivative interactions in $\chi$PT and $\lambda_{\rm eff}$, of Eq.~(\ref{eq:lambdaeff}), is effectively the strength of the repulsive quartic interaction in $\chi$PT.
The resulting dispersion relations for the optical ($\omega\equiv\omega_{o}$) and acoustic ($\omega\equiv\omega_{a}$) modes are
\begin{align}
\label{eq:optical}
&\omega_{o}(\bk,k_{\z})=\mu_{I-}(\bk)\pm\sqrt{k_{\z}^{2}+\mu_{I-}(\bk)^{2}+\tilde{h}_{-}(\bk)}\ ,\\
\label{eq:acoustic}
&\omega_{a}(\bk,k_{\z})=\mu_{I+}(\bk)\pm\sqrt{k_{\z}^{2}+\mu_{I+}(\bk)^{2}+\tilde{h}_{+}(\bk)}\ ,
\end{align}
where $\mu_{I\pm}(\bk)=\mu_{I}[1-\ell^{2}g_{\pm}(\bk)]$. Since $\tilde{h}_{+}(\bzero)=0$ but $\tilde{h}_{-}(\bzero)\neq0$, one of the acoustic modes ($\omega_{a}$) is massless while the remaining acoustic mode and two optical modes are massive. For the massless mode, the dispersion relation for small transverse quasimomentum with $k_{\z}=0$  is
\begin{align}
|\omega_{a}(\bk,0)|&=\tfrac{\alpha_{4}h_{+}(\mathbf{0})}{|\mu_{I+}(\mathbf{0})|}\left(\tfrac{L_{B}\bk}{\pi}\right)^{4}\ ,
\end{align}
where $\alpha_{4}\equiv 2(2q_{4}-\tilde{q}_{4})\approx 5.43$ and $h_{+}(\mathbf{0})$ is defined in Eq.~(\ref{eq:hpm}) with $\Lambda_{+}(\mathbf{0})=2q_{0}\approx1.34$. $q_{i}$ represent coefficients associated with the small quasimomentum expansion of the four-point functions discussed in \ref{sec:appendix}, in particular see Eqs.~(\ref{eq:Q1}) and (\ref{eq:Q2}).

The dispersion relation of the massless mode is softer compared to the corresponding relation in the absence of a chemical potential for which the dispersion relation is quadratic in the infrared limit~\cite{Chernodub:2014sia,rosenstein2010ginzburg}. In the absence of a chemical potential ($\mu_{I}=0$) and derivative interactions ($\ell=0$), our result reduces to that of the Abelian Higgs model (assuming a negative $m_{\pi}^{2}$ necessary for spontaneous symmetry breaking): the massless mode in Eq.~(\ref{eq:optical}) is indeed quadratic for small $\bk$ and $k_{\z}=0$. 

On the left panel of Fig.~\ref{fig:dispersion}, we plot the dispersion relation as a function of the transverse quasimomentum for small values of $\bk$ using $f_{\pi}=93\ {\rm MeV}$ and $m_{\pi}=140\ {\rm MeV}$. We find that for a fixed magnetic field (relative to $B_{c}$) as the isospin chemical potential increases the energy of the massless mode decreases. Similarly, with an increase in the magnetic field (relative to $B_{c}$), the energy decreases. On the right panel of Fig.~\ref{fig:dispersion}, we plot the speed of the massless mode in the transverse direction assuming $k_{\z}=0$. Using the definition of the transverse speed, $\frac{\partial|\omega_{a}(\bk,0)|}{\partial \bk}$, we find that for small $\bk$, up to $\mathcal{O}(|\bk|^{3}|C|^{2})$, it is
\begin{align}
v_{\perp}&=\tfrac{2\alpha_{4}|C|^{2}L_{B}^{4}(\lambda_{\rm eff}-e^{2})}{Q_{0}\pi^{4}|\mu_{I+}(\mathbf{0})|}|\bk|^{3}
\end{align}
with the expression valid for $\frac{L_{B}|\bk|}{\pi}\ll 1$, which is consistent with the LLL approximation for which $|\vec{k}|\ll \sqrt{2eB_{\rm ext}}$. We find that the transverse speed for $\mu_{I}=1.25m_{\pi}$ and $B_{\rm ext}=0.99B_{c}$ is approximately a factor of a thousand smaller than the speed of light for small transverse quasimomentum, $30\lesssim|\mathbf{k}|\lesssim70\ {\rm MeV}$. The speed decreases with increasing magnetic field and isospin chemical potential. Finally, we note that the speed of the massless mode in the longitudinal direction is $\frac{\partial|\omega_{a}(\bzero,k_{\z})|}{\partial k_{\z}}$,
\begin{align}
v_{\parallel}=\tfrac{|k_{\z}|}{\sqrt{k_{\z}^{2}+\mu_{I+}^{2}(\bzero)^{2}}}\ .
\end{align}
In contrast to the AHM, the longitudinal mode does not propagate at the the speed of light~\cite{Chernodub:2014sia} -- it increases monotonically from zero to approximately $0.25$ as the longitudinal momentum $k_{\z}$ is increased from $0\ {\rm MeV}$ to $50\ {\rm MeV}$ and asymptotes to the speed of light for large $k_{\z}$. Similar to the transverse mode, the speed decreases with the increase of the isospin chemical potential and the increase of the magnetic field. However, the effect of the magnetic field is an order of magnitude weaker compared to the transverse mode since the only magnetic field dependence enters through $|\mu_{I+}(\mathbf{0})|$, see Eq.~(\ref{eq:acoustic}), a sub-leading effect.

\section{Conclusion and implications for the phase diagram}
In this paper, we have studied and characterized the dispersion relation of phonon modes associated with the magnetic vortex lattice near the upper critical magnetic field in the LLL approximation, which is valid for small quasimomenta. We find that the energy of the acoustic mode ($\omega_{a}$) is proportional to the fourth power of the transverse quasimomentum, i.e. $|\bk|^{4}$, for small $\bk$, which is softer than the ``supersoft" modes observed in the AHM~\cite{rosenstein2010ginzburg}. The extreme supersoft nature of the phonon mode suggests that the vortex lattice can be easily perturbed away from their mean positions, which raises serious doubts about whether it exists in an ordered lattice phase as has been assumed. 

There are regimes in the standard model, for instance the rho-condensed vacuum of QCD~\cite{Braguta:2012fol} and the $W^{\pm}, Z$ condensed vacuum~\cite{Chernodub:2022ywg} of electroweak theory that form at large magnetic fields in which the vortices exist in as either a ``vortex liquid" or a ``vortex gas". In these examples, the dispersion relations in the long wavelength limit are quadratic and the the interaction energy per unit length between two straight vortices represented by $J_{+}(\mathbf{x})=Q_{+}\delta^{(2)}(\mathbf{x}-\mathbf{x}_{i})$ at $\mathbf{x}_{1}$ and $\mathbf{x}_{2}$ divergences as $\sim-Q_{+}^{2}\log \tfrac{R}{R_{0}}$ with $R=|\mathbf{x}_{1}-\mathbf{x}_{2}|$ for large separations with the negative sign implying repulsive interactions.

Since the dispersion relation in finite isospin $\chi$PT is softer an analogous calculation results in an interaction energy per unit length of $\sim-Q_{+}^{2}M^{2}R^{2}\log \tfrac{R}{R_{0}}$ (where we have introduced a mass $M$ for dimensional consistency), which is significantly more repulsive~\footnote{The resulting interaction is attractive for a vortex, $J_{+}=Q_{+}\delta^{(2)}(\mathbf{x}-\mathbf{x}_{+})$, and an antivortex $J_{-}=Q_{-}\delta^{(2)}(\mathbf{x}-\mathbf{x}_{-})$, with the energy per unit length scaling as $+Q_{+}Q_{-}M^{2}R^{2}\log \tfrac{R}{R_{0}}$ where $R=|\mathbf{x}_{+}-\mathbf{x}_{-}|$. Additionally, in the infrared, the interaction is more strongly confining than the linear potential of QCD~\cite{creutz1980monte,creutz1982numerical}, dual QCD~\cite{baker1991dual} or that observed in $SU(2)$ non-Abelian Higgs theory with strong magnetic fields.~\cite{suganuma2018non}}. As such, it seems reasonable to expect the magnetic vortex lattice of isospin $\chi$PT near the critical magnetic field to exist in a spaghetti phase of either a  ``liquid" or ``gas" phase as opposed to an ordered lattice.

\begin{figure}
\begin{center}
\includegraphics[width=0.8\textwidth]{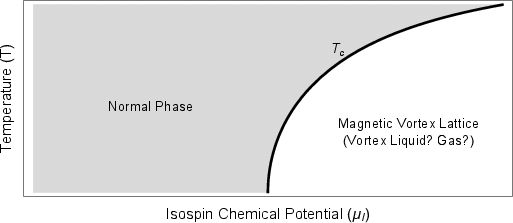}
\end{center}
\caption{Schematic, finite temperature phase diagram: at the critical temperature, $T_{c}$, the vortex lattice is destroyed by thermal fluctuations.}
\label{fig:phaseT}
\end{figure}

Unfortunately, lattice studies are not possible for finite isospin QCD in a magnetic background (with dynamical gauge fields) due to the fermion sign problem. However, it is possible to incorporate quantum effects using techniques similar to Ref~\cite{Haber:2017kth} (where an ordered lattice was assumed): thermal fluctuations destroy the ordered magnetic vortex lattice in $\chi$PT at a critical temperature of $T_{c}\sim f_{\pi}\sqrt{1-m_{\pi}^{2}/\mu_{I}^{2}}$~\cite{Haber:2017kth}, which we schematically illustrate  Fig.~\ref{fig:phaseT} with  temperature and finite isospin chemical potentials plotted on the axes. Above the critical temperature $T_{c}$, the normal vacuum with no pion condensate is favored. Below this temperature, our results suggest the possibility of magnetic vortices existing in a disordered spaghetti phase of either a vortex liquid or gas. If in a disordered phase, there may be further finite temperature phase transitions in the phase diagram. Furthermore, there may be phenomenological consequences associated with this new phase -- pion stars have been conjectured to form in the remnants of supernova explosion~\cite{Carignano:2016lxe} with $\chi$PT predicting their mass-radius relationship~\cite{brandt2018new}. In the presence of weak magnetic fields, the equation of state is altered, which in turn affects the size of magnetized stars.

We conclude by noting that in the ordered phase, the phase diagram in the magnetic field-isospin chemical potential plane, is richer than traditionally depicted. In particular, we note the symmetries of a uniform superconductor, a single magnetic vortex and multiple vortices (including the vortex lattice) 
\begin{align}
\textrm{Superconductor: } &\mathbb{R}_{3}\times SO(3)\cong E(3) \\
\nonumber&\downarrow \scalebox{0.7}{$H\ge H_{c1}$}\\
\textrm{Single Vortex:     } &\mathbb{R}_{\z}\times SO(2)_{\theta+\alpha}\\
\nonumber&\downarrow \scalebox{0.8}{$H\ge H_{c3}$}\\
\textrm{Multiple Vortices: }&\mathbb{R}_{\z}\times \emptyset\ ,
\end{align}
where $R_{3}$ and $SO(3)$ represent translations and rotations (respectively) in three-dimensions, $R_{\z}$ represents translation in the $\z$-direction and $SO(2)_{\theta+\alpha}$ represents simultaneous physical and phase rotations of a single vortex, with the system being invariant if the angles are equal. In the thermodynamic limit, $H_{c3}$ approaches $H_{c1}$ from above -- as such the phase transition line at $H_{c1}$ supports not only a single vortex phase but also a distinct phase of a finite density of magnetic vortices with the phase sustained for external magnetic fields above $H_{c1}$ and below $H_{c2}$.
\label{sec:conclusion}
\section*{Acknowledgements}
\noindent
P.A. would also like to thank Maxim Chernodub for helpful correspondence on quasimomentum states, and Andreas Schmitt for not only bringing Ref.~\cite{Haber:2017kth} to our attention but also for a discussion of the critical temperature $T_{c}$ in finite isospin $\chi$PT. P.A. would also like to acknowledge Patrick Jefferson for helpful discussions on related issues. E.L. and J.M. would like to acknowledge the Collaborative Undergraduate Research and Inquiry (CURI) and Directed Undergraduate Research (DUR) programs for their support. Finally, P.A. acknowledges the support of St. Olaf startup funds and the support of the Kavli Institute for Theoretical Physics, Santa Barbara, through which the research was supported in part by the National Science Foundation under Grant No. NSF PHY-1748958.
\appendix
\section{Useful Mathematical Results: Four-Point Function}
\label{sec:appendix}
\noindent
For notational convenience, we use the dimensionless quasimomentum $\overline{\bk}$ and position $\overline{\mathbf{x}}$, which are
\begin{align}
\overline{\bk}=\frac{L_{B}}{\pi}\bk\ ,\ \overline{\mathbf{x}}=\frac{1}{L_{B}}\mathbf{x}\ ,
\end{align}
where $L_{B}=\sqrt{\frac{2\pi}{eB_{\rm ext}}}$ is the magnetic length. In terms of the dimensionless $\overline{\mathbf{x}}$ and $\overline{\bk}$,
the four-point function is defined as the transverse average -- Eq.~(\ref{eq:Q}) -- of the product of four pion fields with quasi-momenta,
\begin{align}
\label{eq:4-point-function}
\langle\tilde{\pi}^{+}_{\overline{\mathbf{l}}_{2}}\tilde{\pi}^{-}_{\overline{\bk}_{2}}\tilde{\pi}^{+}_{\overline{\mathbf{l}}_{1}}\tilde{\pi}^{-}_{\overline{\bk}_{1}}\rangle&\equiv|C|^{4}Q_{\overline{\mathbf{l}}_{2}\overline{\bk}_{2}\overline{\mathbf{l}}_{1}\overline{\bk}_{1}}\ ,
\end{align}
where $|C|$ has mass dimension one and $Q_{\overline{\mathbf{l}}_{2}\overline{\bk}_{2}\overline{\mathbf{l}}_{1}\overline{\bk}_{1}}$ is dimensionless,
\begin{align}
Q_{\overline{\mathbf{l}}_{2}\overline{\bk}_{2}\overline{\mathbf{l}}_{1}\overline{\bk}_{1}}=Q_{0}^{2}\sum_{m,n\in \mathbb{Z}}e^{-\pi(\mathbf{c}-\mathbf{X}_{m,n})^{2}+2\pi i\epsilon_{ij}b^{i}X^{j}_{m,n}}
\end{align}
with $\mathbf{b}=\frac{\overline{\mathbf{k}}_{1}+\overline{\mathbf{l}}_{1}-\overline{\bk}_{2}-\overline{\mathbf{l}}_{2}}{2}$ and $\mathbf{c}=\frac{\overline{\mathbf{l}}_{1}-\overline{\bk}_{1}-\overline{\mathbf{l}}_{2}+\overline{\bk}_{2}}{2}$.
$\mathbf{X}_{m,n}$ is defined in terms of the lattice vectors $\mathbf{d}_{1}$ and $\mathbf{d}_{2}$ as
\begin{align}
\mathbf{X}_{m,n}&=\left(m\tilde{\mathbf{d}}_{2}+n\mathbf{\tilde{d}}_{1}\right)=\tfrac{(m+n)}{2\nu}\mathbf{e}_{x}-m\nu\mathbf{e}_{y}
\end{align}
where $\tilde{\mathbf{d}}_{a}=\frac{1}{L_{B}}\epsilon_{ab}\mathbf{d}_{b}$, and $\mathbf{e}_{x}$ and $\mathbf{e}_{y}$ are unit vectors. For a detailed derivation of the four-point function in the symmetric gauge, see Ref.~\cite{Chernodub:2014sia}.  The lattice vectors associated with the triangular magnetic vortex lattice is
\begin{align}
\mathbf{d}_{1}&=\tfrac{L_{B}}{\nu}\mathbf{e}_{y}\ ,\ \mathbf{d}_{2}=\tfrac{L_{B}}{\nu}\left(\tfrac{\sqrt{3}}{2}\mathbf{e}_{x}+\tfrac{1}{2}\mathbf{e}_{y}\right)
\end{align}
where $\nu=\tfrac{1}{\sqrt{2}Q_{0}}=\frac{\sqrt[4]{3}}{\sqrt{2}}$ for the triangular lattice. The four-point function is symmetric under the following non-trivial exchange of quasimomentum indices.
\begin{align}
Q_{\overline{\mathbf{l}}_{2}\overline{\bk}_{2}\overline{\mathbf{l}}_{1}\overline{\bk}_{1}}=Q_{\overline{\bk}_{2}\overline{\mathbf{l}}_{2}\overline{\bk}_{1}\overline{\mathbf{l}}_{1}}\ .
\end{align} 
This is in addition to trivial exchanges allowed by Eq.~(\ref{eq:Q}). We require the following small $\bk$ expansion of $Q_{\bk\bk\mathbf{0}\mathbf{0}}$, $Q_{\bk\bzero\bzero\bk}$ and $Q_{\bk\bzero-\bk\bzero}$ with two non-zero momenta.
\begin{align}
\label{eq:Q1}
&Q_{\bk\bk\bzero\bzero}=q_{0}+q_{2}\left(\tfrac{L_{B}}{\pi}\bk\right)^{2}+q_{4}\left(\tfrac{L_{B}}{\pi}\bk\right)^{4}+\cdots\\
\label{eq:Q2}
&Q_{\bk\bzero-\bk\bzero}=q_{0}+\tilde{q}_{2}\left(\tfrac{L_{B}}{\pi}\bk\right)^{2}+\tilde{q}_{4}\left(\tfrac{L_{B}}{\pi}\bk\right)^{4}+\cdots
\end{align}
where $Q_{\bk\bk\bzero\bzero}=Q_{\bk\bzero\bzero\bk}$ due to Eq.~(\ref{eq:4-point-function}) and the coefficients are
\begin{align}
q_{0}&=\tfrac{\beta_{A}}{2\nu^{2}}=0.669493\dots\\
q_{2}&=-1.05164\dots\ ,\ q_{4}=3.00957\dots\\
\tilde{q}_{2}&=-2.10327\dots\ ,\ \tilde{q}_{4}=3.30381\dots
\end{align}
with $\tilde{q}_{2}=2q_{2}$. As noted in the derivation of Ref.~\cite{Chernodub:2014sia}, the expression for the four-point correlation function holds only for small quasimomenta, $|\overline{k}_{x}|\le\frac{\sqrt{2}Q_{0}}{4}$ and $|\overline{k}_{y}|\le\frac{1}{2\sqrt{2}Q_{0}}$ with $Q_{0}=\frac{1}{\sqrt{2}\nu}$ and $\nu=\frac{\sqrt[4]{3}}{\sqrt{2}}$ for a triangular vortex lattice.
\section{Derivation of Quadratic Phonon Action}
\label{app:phononaction}
\subsection{Transverse Energy}
\label{app:transverse}
\noindent
The transverse energy that incorporates quadratic fluctuations can be found using Eq.~(\ref{eq:E}) after making the replacements in Eqs.~(\ref{eq:rep1}) and (\ref{eq:rep2}),
\begin{align}
\nonumber
\mathcal{E}_{\perp}^{(2)}&=-e(B_{c}-B_{\rm ext})\langle\chi^{\dagger}\chi\rangle+e^{2}\langle\chi^{\dagger}\chi\rangle\langle\tilde{\pi}_{0}^{+}\tilde{\pi}_{0}^{-}\rangle\\
\nonumber
&+\tfrac{1}{2}(\lambda_{\rm eff}-e^{2})\left[\langle\chi^{\dagger}\chi\tilde{\pi}_{0}^{+}\tilde{\pi}_{0}^{-}\rangle+\langle\tilde{\pi}_{0}^{+}\tilde{\pi}_{0}^{-}\chi^{\dagger}\chi\rangle+\langle\chi^{\dagger}\tilde{\pi}_{0}^{-}\tilde{\pi}_{0}^{+}\chi\rangle+\langle\tilde{\pi}_{0}^{+}\chi\chi^{\dagger}\tilde{\pi}_{0}^{-}\rangle\right.\\
&\left.+\langle\chi^{\dagger}\tilde{\pi}^{-}_{0}\chi^{\dagger}\tilde{\pi}^{-}_{0}\rangle+\langle\tilde{\pi}^{+}_{0}\chi\tilde{\pi}^{+}_{0}\chi\rangle\right]\ ,
\end{align}
where we have used the orthogonality of the fluctuations to the static solution, $\langle\tilde{\pi}^{-}_{0}\chi\rangle=\langle\chi^{\dagger}\tilde{\pi}_{0}^{+}\rangle=0$. Since $\mathcal{E}_{\perp}$ is strictly valid for static solutions with $k_{\z}=0$, we replace $\chi$ using Eq.~(\ref{eq:chi}) after setting $k_{\z}=0$. Using Eq.~(\ref{eq:Qproperty}) and quasimomentum conservation implicit in the four-point function~\cite{Chernodub:2014sia}, we find
\begin{align}
\mathcal{E}^{(2)}_{\perp}&=\sum_{\bk\neq\bzero}\mathcal{E}^{(2)}_{\perp}(\bk,0)\\
\nonumber
\mathcal{E}_{\perp}^{(2)}(\bk,0)&=\left[-e(B_{c}-B_{\rm ext})|C|^{2}Q_{0}+e^{2}|C|^{4}Q_{0}^{2}\right.\\
&\left.+(\lambda_{\rm eff}-e^{2})|C|^{4}(Q_{\bk\bk\bzero\bzero}+Q_{\bk\bzero\bzero\bk})\right]c^{\dagger}_{\bk}c_{\bk}\\
\nonumber
&+\left[(\lambda_{\rm eff}-e^{2})|C|^{4}Q_{\bk\bzero-\bk\bzero}\right] \tfrac{1}{2}(c^{\dagger}_{\bk}c^{\dagger}_{-\bk}+c_{\bk}c_{-\bk})\ ,
\end{align}
where $Q_{0}\equiv\frac{1}{\sqrt{2}\nu}$. 
The diagonal form of $\mathcal{E}_{\perp}^{(2)}(\bk,0)$ is
\begin{align}
\nonumber
\mathcal{E}_{\perp}^{(2)}(\bk,0)=&\tfrac{1}{2}(\lambda_{\rm eff}-e^{2})|C|^{4}\left[\lambda_{-}(\bk)-\lambda_{-}(\bzero)\right]o_{\bk}^{2}\\
\label{eq:Eperpk}
+&\tfrac{1}{2}(\lambda_{\rm eff}-e^{2})|C|^{4}\left[\lambda_{+}(\bk)-\lambda_{-}(\bzero)\right]a_{\bk}^{2}
\end{align}
where $\lambda_{\pm}(\bk)$ is defined as
\begin{align}
\label{eq:lambda}
\lambda_{\pm}(\bk)&\equiv2(Q_{\bk\bk\bzero\bzero}+Q_{\bk\bzero\bzero\bk}\pm Q_{\bk\bzero-\bk\bzero})\ ,
\end{align}
and $\lambda_{-}(\bzero)=2Q_{0}^{2}\beta_{A}$ in conjunction with Eq.~(\ref{eq:C}) gives the compact form of $\mathcal{E}_{\perp}^{(2)}(\bk,0)$ in Eq.~(\ref{eq:Eperpk}). 

\subsection{Longitudinal Energy}
\label{app:longitudinal}
\noindent 
The contribution to the quadratic phonon Lagrangian of the longitudinal quasimomentum is most conveniently calculated using the derivative terms in Eq.~(\ref{eq:L4}) that depend on the longitudinal direction. We get
\begin{align}
\mathcal{E}_{\parallel}^{(2)}=\mathcal{E}_{\parallel,0}^{(2)}+\mathcal{E}_{\parallel,d}^{(2)}=\sum_{k\neq 0}\left(\mathcal{E}_{\parallel,0}^{(2)}(k)+\mathcal{E}_{\parallel,d}^{(2)}(k)\right)\ ,
\end{align}
where the contribution due to the leading derivative term is $\mathcal{E}_{\parallel,0}^{(2)}=\langle\partial_{\z}\chi^{\dagger}\partial_{\z}\chi\rangle$, which gives
\begin{align}
\mathcal{E}_{\parallel,0}^{(2)}(k)=k_{\z}^{2}|C|^{2}Q_{0}c^{\dagger}_{k}c_{k}\ .
\end{align}
The sub-leading derivative terms in $\mathcal{L}_{4\pi}$ contribute to
\begin{align}
\nonumber
\mathcal{E}_{\parallel,d}^{(2)}(k)&=-k_{\z}^{2}|C|^{4}\ell^{2}\left[Q_{\bk\bk\bzero\bzero}c^{\dagger}_{k}c_{k}+Q_{\bk\bzero-\bk\bzero}\tfrac{1}{2}(c^{\dagger}_{k}c^{\dagger}_{-k}+c_{k}c_{-k})\right]\ ,
\end{align}
where $\ell\equiv\frac{1}{\sqrt{3}f_{\pi}}$.
In the basis of the real optical and acoustic ``operators" $o_{k}$ and $a_{k}$,
\begin{align}
\mathcal{E}_{\parallel}^{(2)}(k)=&k_{\z}^{2}|C|^{2}\left[Q_{0}-|C|^{2}\ell^{2}\Lambda_{+}(\bk)\right]o_{k}^{2}+k_{\z}^{2}|C|^{2}\left[Q_{0}-|C|^{2}\ell^{2}\Lambda_{-}(\bk)\right]a_{k}^{2}\ ,
\end{align}
where $\Lambda_{\pm}(\bk)=Q_{\bk\bk00}\pm Q_{\bk0-\bk0}$.

\subsection{Kinetic Energy}
\label{app:kinetic}
\noindent

The calculation of the time-dependent contribution of the quadratic phonon Lagrangian proceeds in a manner similar to the longitudinal ($k_{\z}$-dependent) contribution.  However, as is evident from Eq.~(\ref{eq:L4}), there are also single time-derivative contributions proportional to the finite isospin chemical potential. We find
\begin{align}
\nonumber
\mathcal{T}^{(2)}&=\mathcal{T}_{0}^{(2)}+\mathcal{T}_{d}^{(2)}=\sum_{k\neq 0}\left(\mathcal{T}_{0}^{(2)}(k)+\mathcal{T}_{d}^{(2)}(k)\right).
\end{align}
where $\mathcal{T}_{0}^{(2)}$ is the contribution arising from the $\mathcal{O}(p^{2})$ Lagrangian 
\begin{align}
\mathcal{T}_{0}^{(2)}&=\langle\partial_{0}\chi^{\dagger}\partial^{0}\chi+i\mu_{I}(\partial_{0}\chi^{\dagger}\chi-\chi^{\dagger}\partial_{0}\chi)\rangle\ .
\end{align}
Upon utilization of the quasimomentum expansion for $\chi$, we find that
\begin{align}
\mathcal{T}^{(2)}_{0}(k)=\left[\omega^{2}-2\mu_{I}\omega\right]|C|^{2}Q_{0}c^{\dagger}_{k}c_{k}\ .
\end{align}
$\mathcal{T}_{d}^{(2)}$ is the contribution arising from the $\mathcal{O}(p^{4})$ derivative terms, 
\begin{align}
\nonumber
\mathcal{T}_{d}^{(2)}=&-2i\mu_{I}\ell^{2}\left[\langle\tilde{\pi}^{+}_{0}\tilde{\pi}^{-}_{0}(\partial_{0}\chi^{\dagger}\chi-\chi^{\dagger}\partial_{0}\chi)\rangle\right.\\
\nonumber
&+\langle\tilde{\pi}^{+}_{0}\chi(\partial_{0}\chi^{\dagger}\tilde{\pi}^{-}_{0}-\tilde{\pi}^{+}_{0}\partial_{0}\chi)\rangle\left.+\chi^{\dagger}\tilde{\pi}^{-}_{0}(\partial_{0}\chi^{\dagger}\tilde{\pi}^{-}_{0}-\tilde{\pi}^{+}_{0}\partial_{0}\chi)\rangle\right]\\
\nonumber
&-\ell^{2}\left[\tilde{\pi}^{+}_{0}\tilde{\pi}^{-}_{0}\partial_{0}\chi^{\dagger}\partial_{0}\chi-\tfrac{1}{2}(\tilde{\pi}^{+}_{0}\partial_{0}\chi\tilde{\pi}^{+}_{0}\partial_{0}\chi+\partial_{0}\chi^{\dagger}\tilde{\pi}^{-}_{0}\partial_{0}\chi^{\dagger}\tilde{\pi}^{-}_{0})\right]\ .
\end{align}
Utilizing the quasimomentum expansion of $\chi$, we find that
\begin{align}
\nonumber
\mathcal{T}_{d}^{(2)}(k)&=4\ell^{2}\mu_{I}\omega|C|^{4}\left[(Q_{\bk\bk\bzero\bzero}+Q_{\bk\bzero\bzero\bk})c^{\dagger}_{k}c_{k}+Q_{\bk\bzero-\bk\bzero}\tfrac{1}{2}(c^{\dagger}_{k}c^{\dagger}_{-k}+c_{k}c_{-k})\right]\\
&-\omega^{2}\ell^{2}|C|^{4}\left[Q_{\bk\bk\bzero\bzero}c^{\dagger}_{k}c_{k}+Q_{\bk\bzero-\bk\bzero}\tfrac{1}{2}(c^{\dagger}_{k}c^{\dagger}_{-k}+c_{k}c_{-k})\right]\ ,
\end{align}
which in terms of the optical and acoustic ``operators", $o_{k}$ and $a_{k}$, respectively are
\begin{align}
\nonumber
\mathcal{T}^{(2)}=&(\omega^{2}-2\mu_{I}\omega)|C|^{2}\left[Q_{0}-|C|^{2}\ell^{2}\Lambda_{+}(\bk)\right]o_{k}^{2}\\
\nonumber
+&(\omega^{2}-2\mu_{I}\omega)|C|^{2}\left[Q_{0}-|C|^{2}\ell^{2}\Lambda_{-}(\bk)\right]a_{k}^{2}\\
+&2\ell^{2}\mu_{I}\omega|C|^{4}(\tilde{\Lambda}_{+}(\bk)o_{k}^{2}+\tilde{\Lambda}_{-}(\bk)a_{k}^{2})\ ,
\end{align}
where $\tilde{\Lambda}_{\pm}(\bk)=Q_{\bk\bk\bzero\bzero}+2Q_{\bk\bzero\bzero\bk}\pm Q_{\bk\bzero-\bk\bzero}$.

\bibliographystyle{elsarticle-num} 
\bibliography{bib}

\begin{thebibliography}{10}
\expandafter\ifx\csname url\endcsname\relax
  \def\url#1{\texttt{#1}}\fi
\expandafter\ifx\csname urlprefix\endcsname\relax\def\urlprefix{URL }\fi
\expandafter\ifx\csname href\endcsname\relax
  \def\href#1#2{#2} \def\path#1{#1}\fi

\bibitem{Son:2000xc}
D.~T. Son, M.~A. Stephanov, {QCD at finite isospin density}, Phys. Rev. Lett.
  86 (2001) 592--595.
\newblock \href {http://arxiv.org/abs/hep-ph/0005225}
  {\path{arXiv:hep-ph/0005225}}, \href
  {https://doi.org/10.1103/PhysRevLett.86.592}
  {\path{doi:10.1103/PhysRevLett.86.592}}.

\bibitem{Son:2000by}
D.~T. Son, M.~A. Stephanov, {QCD at finite isospin density: From pion to quark
  - anti-quark condensation}, Phys. Atom. Nucl. 64 (2001) 834--842, [Yad.
  Fiz.64,899(2001)].
\newblock \href {http://arxiv.org/abs/hep-ph/0011365}
  {\path{arXiv:hep-ph/0011365}}, \href {https://doi.org/10.1134/1.1378872}
  {\path{doi:10.1134/1.1378872}}.

\bibitem{Vovchenko:2020crk}
V.~Vovchenko, B.~B. Brandt, F.~Cuteri, G.~Endr\H{o}di, F.~Hajkarim,
  J.~Schaffner-Bielich, {Pion Condensation in the Early Universe at
  Nonvanishing Lepton Flavor Asymmetry and Its Gravitational Wave Signatures},
  Phys. Rev. Lett. 126~(1) (2021) 012701.
\newblock \href {http://arxiv.org/abs/2009.02309} {\path{arXiv:2009.02309}},
  \href {https://doi.org/10.1103/PhysRevLett.126.012701}
  {\path{doi:10.1103/PhysRevLett.126.012701}}.

\bibitem{Carignano:2016lxe}
S.~Carignano, L.~Lepori, A.~Mammarella, M.~Mannarelli, G.~Pagliaroli,
  {Scrutinizing the pion condensed phase}, Eur. Phys. J. A53~(2) (2017) 35.
\newblock \href {http://arxiv.org/abs/1610.06097} {\path{arXiv:1610.06097}},
  \href {https://doi.org/10.1140/epja/i2017-12221-x}
  {\path{doi:10.1140/epja/i2017-12221-x}}.

\bibitem{brandt2018new}
B.~B. Brandt, G.~Endr{\H{o}}di, E.~S. Fraga, M.~Hippert, J.~Schaffner-Bielich,
  S.~Schmalzbauer, New class of compact stars: Pion stars, Physical Review D
  98~(9) (2018) 094510.

\bibitem{Andersen:2018nzq}
J.~O. Andersen, P.~Kneschke, {Bose-Einstein condensation and pion stars} (7
  2018).
\newblock \href {http://arxiv.org/abs/1807.08951} {\path{arXiv:1807.08951}}.

\bibitem{Abuki:2009hx}
H.~Abuki, T.~Brauner, H.~J. Warringa, {Pion condensation in a dense neutrino
  gas}, Eur. Phys. J. C 64 (2009) 123--131.
\newblock \href {http://arxiv.org/abs/0901.2477} {\path{arXiv:0901.2477}},
  \href {https://doi.org/10.1140/epjc/s10052-009-1121-0}
  {\path{doi:10.1140/epjc/s10052-009-1121-0}}.

\bibitem{Dashen:1974pio}
R.~Dashen, J.~Manassah,
  \href{https://www.sciencedirect.com/science/article/pii/0375960174905738}{Pion
  phase transition and phonon excitation spectrum in chiral symmetry breaking
  model}, Physics Letters A 47~(6) (1974) 453--454.
\newblock \href {https://doi.org/https://doi.org/10.1016/0375-9601(74)90573-8}
  {\path{doi:https://doi.org/10.1016/0375-9601(74)90573-8}}.
\newline\urlprefix\url{https://www.sciencedirect.com/science/article/pii/0375960174905738}

\bibitem{Kanazawa:2011tt}
T.~Kanazawa, T.~Wettig, N.~Yamamoto, {Singular values of the Dirac operator in
  dense QCD-like theories}, JHEP 12 (2011) 007.
\newblock \href {http://arxiv.org/abs/1110.5858} {\path{arXiv:1110.5858}},
  \href {https://doi.org/10.1007/JHEP12(2011)007}
  {\path{doi:10.1007/JHEP12(2011)007}}.

\bibitem{Kanazawa:2014lga}
T.~Kanazawa, T.~Wettig, {Stressed Cooper pairing in QCD at high isospin
  density: effective Lagrangian and random matrix theory}, JHEP 10 (2014) 055.
\newblock \href {http://arxiv.org/abs/1406.6131} {\path{arXiv:1406.6131}},
  \href {https://doi.org/10.1007/JHEP10(2014)055}
  {\path{doi:10.1007/JHEP10(2014)055}}.

\bibitem{Cohen:2015soa}
T.~D. Cohen, S.~Sen, {Deconfinement Transition at High Isospin Chemical
  Potential and Low Temperature}, Nucl. Phys. A942 (2015) 39--53.
\newblock \href {http://arxiv.org/abs/1503.00006} {\path{arXiv:1503.00006}},
  \href {https://doi.org/10.1016/j.nuclphysa.2015.07.018}
  {\path{doi:10.1016/j.nuclphysa.2015.07.018}}.

\bibitem{Kogut:2002zg}
J.~B. Kogut, D.~K. Sinclair, {Lattice QCD at finite isospin density at zero and
  finite temperature}, Phys. Rev. D 66 (2002) 034505.
\newblock \href {http://arxiv.org/abs/hep-lat/0202028}
  {\path{arXiv:hep-lat/0202028}}, \href
  {https://doi.org/10.1103/PhysRevD.66.034505}
  {\path{doi:10.1103/PhysRevD.66.034505}}.

\bibitem{Brandt:2017zck}
B.~B. Brandt, G.~Endrodi, S.~Schmalzbauer, {QCD at finite isospin chemical
  potential}, EPJ Web Conf. 175 (2018) 07020.
\newblock \href {http://arxiv.org/abs/1709.10487} {\path{arXiv:1709.10487}},
  \href {https://doi.org/10.1051/epjconf/201817507020}
  {\path{doi:10.1051/epjconf/201817507020}}.

\bibitem{Mannarelli:2019hgn}
M.~Mannarelli, {Meson condensation}, Particles 2~(3) (2019) 411--443.
\newblock \href {http://arxiv.org/abs/1908.02042} {\path{arXiv:1908.02042}},
  \href {https://doi.org/10.3390/particles2030025}
  {\path{doi:10.3390/particles2030025}}.

\bibitem{He:2005sp}
L.~He, P.~Zhuang, {Phase structure of Nambu-Jona-Lasinio model at finite
  isospin density}, Phys. Lett. B615 (2005) 93--101.
\newblock \href {http://arxiv.org/abs/hep-ph/0501024}
  {\path{arXiv:hep-ph/0501024}}, \href
  {https://doi.org/10.1016/j.physletb.2005.03.066}
  {\path{doi:10.1016/j.physletb.2005.03.066}}.

\bibitem{Adhikari:2018cea}
P.~Adhikari, J.~O. Andersen, P.~Kneschke, {Pion condensation and phase diagram
  in the Polyakov-loop quark-meson model}, Phys. Rev. D98~(7) (2018) 074016.
\newblock \href {http://arxiv.org/abs/1805.08599} {\path{arXiv:1805.08599}},
  \href {https://doi.org/10.1103/PhysRevD.98.074016}
  {\path{doi:10.1103/PhysRevD.98.074016}}.

\bibitem{Adhikari:2015wva}
P.~Adhikari, T.~D. Cohen, J.~Sakowitz, {Finite Isospin Chiral Perturbation
  Theory in a Magnetic Field}, Phys. Rev. C91~(4) (2015) 045202.
\newblock \href {http://arxiv.org/abs/1501.02737} {\path{arXiv:1501.02737}},
  \href {https://doi.org/10.1103/PhysRevC.91.045202}
  {\path{doi:10.1103/PhysRevC.91.045202}}.

\bibitem{Adhikari:2018fwm}
P.~Adhikari, {Magnetic Vortex Lattices in Finite Isospin Chiral Perturbation
  Theory}, Phys. Lett. B790 (2019) 211--217.
\newblock \href {http://arxiv.org/abs/1810.03663} {\path{arXiv:1810.03663}},
  \href {https://doi.org/10.1016/j.physletb.2019.01.027}
  {\path{doi:10.1016/j.physletb.2019.01.027}}.

\bibitem{Brauner:2016pko}
T.~Brauner, N.~Yamamoto, {Chiral Soliton Lattice and Charged Pion Condensation
  in Strong Magnetic Fields}, JHEP 04 (2017) 132.
\newblock \href {http://arxiv.org/abs/1609.05213} {\path{arXiv:1609.05213}},
  \href {https://doi.org/10.1007/JHEP04(2017)132}
  {\path{doi:10.1007/JHEP04(2017)132}}.

\bibitem{Brauner:2021sci}
T.~Brauner, H.~Kole\v{s}ov\'a, N.~Yamamoto, {Chiral soliton lattice phase in
  warm QCD}, Phys. Lett. B 823 (2021) 136767.
\newblock \href {http://arxiv.org/abs/2108.10044} {\path{arXiv:2108.10044}},
  \href {https://doi.org/10.1016/j.physletb.2021.136767}
  {\path{doi:10.1016/j.physletb.2021.136767}}.

\bibitem{gronli2022competition}
M.~S. Gr{\o}nli, T.~Brauner, Competition of chiral soliton lattice and
  abrikosov vortex lattice in qcd with isospin chemical potential, The European
  Physical Journal C 82~(4) (2022) 354.

\bibitem{Evans:2022hwr}
G.~W. Evans, A.~Schmitt, {Chiral anomaly induces superconducting baryon
  crystal}, JHEP 09 (2022) 192.
\newblock \href {http://arxiv.org/abs/2206.01227} {\path{arXiv:2206.01227}},
  \href {https://doi.org/10.1007/JHEP09(2022)192}
  {\path{doi:10.1007/JHEP09(2022)192}}.

\bibitem{Abrikosov:1956sx}
A.~A. Abrikosov, {On the Magnetic properties of superconductors of the second
  group}, Sov. Phys. JETP 5 (1957) 1174--1182, [Zh. Eksp. Teor.
  Fiz.32,1442(1957)].

\bibitem{Chernodub:2014sia}
M.~N. Chernodub, J.~Van~Doorsselaere, H.~Verschelde, {Phonon spectrum of the
  QCD vacuum in a magnetic-field-induced superconducting phase}, Phys. Rev.
  D89~(10) (2014) 105011.
\newblock \href {http://arxiv.org/abs/1401.0264} {\path{arXiv:1401.0264}},
  \href {https://doi.org/10.1103/PhysRevD.89.105011}
  {\path{doi:10.1103/PhysRevD.89.105011}}.

\bibitem{Rosenstein:1999aa}
B.~Rosenstein, First-principles theory of fluctuations in vortex liquids and
  solids, Physical Review B 60~(6) (1999) 4268--4271.
\newblock \href {https://doi.org/10.1103/PhysRevB.60.4268}
  {\path{doi:10.1103/PhysRevB.60.4268}}.

\bibitem{rosenstein2010ginzburg}
B.~Rosenstein, D.~Li, Ginzburg-landau theory of type ii superconductors in
  magnetic field, Rev. Mod. Phys. 82~(1) (2010) 109.

\bibitem{Braguta:2012fol}
V.~V. Braguta, P.~V. Buividovich, M.~Chernodub, M.~I. Polikarpov, A.~Y. Kotov,
  {Vortex liquid in magnetic-field-induced superconducting vacuum of quenched
  lattice QCD}, PoS ConfinementX (2012) 083.
\newblock \href {http://arxiv.org/abs/1301.6590} {\path{arXiv:1301.6590}},
  \href {https://doi.org/10.22323/1.171.0083} {\path{doi:10.22323/1.171.0083}}.

\bibitem{Chernodub:2022ywg}
M.~N. Chernodub, V.~A. Goy, A.~V. Molochkov, {Phase Structure of Electroweak
  Vacuum in a Strong Magnetic Field: The Lattice Results}, Phys. Rev. Lett.
  130~(11) (2023) 111802.
\newblock \href {http://arxiv.org/abs/2206.14008} {\path{arXiv:2206.14008}},
  \href {https://doi.org/10.1103/PhysRevLett.130.111802}
  {\path{doi:10.1103/PhysRevLett.130.111802}}.

\bibitem{creutz1980monte}
M.~Creutz, Monte carlo study of quantized su (2) gauge theory, Physical Review
  D 21~(8) (1980) 2308.

\bibitem{creutz1982numerical}
M.~Creutz, K.~Moriarty, Numerical studies of wilson loops in su (3) gauge
  theory in four dimensions, Physical Review D 26~(8) (1982) 2166.

\bibitem{baker1991dual}
M.~Baker, J.~S. Ball, F.~Zachariasen, Dual qcd: a review, Physics Reports
  209~(3) (1991) 73--127.

\bibitem{suganuma2018non}
H.~Suganuma, Non-abelian higgs theory in a strong magnetic field and
  confinement, arXiv preprint arXiv:1812.06843 (2018).

\bibitem{Haber:2017kth}
A.~Haber, A.~Schmitt, {Critical magnetic fields in a superconductor coupled to
  a superfluid}, Phys. Rev. D95~(11) (2017) 116016.
\newblock \href {http://arxiv.org/abs/1704.01575} {\path{arXiv:1704.01575}},
  \href {https://doi.org/10.1103/PhysRevD.95.116016}
  {\path{doi:10.1103/PhysRevD.95.116016}}.

\end{thebibliography}
\end{document}